# 3D Gaze Vis: Sharing Eye Tracking Data Visualization for Collaborative Work in VR Environment

Song Zhao[1], Shiwei Cheng[1], and Chenshuang Zhu[1]

[1] Zhejiang University of Technology, Hangzhou 310023, China
`swc@zjut.edu.cn`

**Abstract.** Conducting collaborative tasks, *e.g.*, multi-user game, in virtual reality (VR) could enable us to explore more immersive and effective experience. However, for current VR systems, users cannot communicate properly with each other via their gaze points, and this would interfere with users' mutual understanding of the intention. In this study, we aimed to find the optimal eye tracking data visualization , which minimized the cognitive interference and improved the understanding of the visual attention and intention between users. We designed three different eye tracking data visualizations: gaze cursor, gaze spotlight and gaze trajectory in VR scene for a course of human heart , and found that gaze cursor from doctors could help students learn complex 3D heart models more effectively. To further explore, two students as a pair were asked to finish a quiz in VR environment, with sharing gaze cursors with each other, and obtained more efficiency and scores. It indicated that sharing eye tracking data visualization could improve the quality and efficiency of collaborative work in the VR environment.

**Keywords:** Gaze fixation, computer supported collaborative learning, information visualization, medical visualization.

## 1 Introduction

Virtual reality (VR) technology provides users with extraordinary immersive entertainment. Software and hardware developers have also made a lot of efforts in terms of experience, for example, adding auditory, haptic and visual approaches to improve the fun in games, and realizing two-player or even multi-player online VR modes  to improve communication between players. However, a major challenge still remains: how to let users collaborate with each other naturally and conveniently as they do in their daily life.

In daily life, people collaborate with each other in many ways, among which  eye contact has been one of the most natural and effective ways. In this way,, people can easily understand which region and object others are currently focusing on. However, only few existing VR collaboration studies use eye tracking as a collaboration technique. One of the major problems in VR is that users cannot communicate through eye contact  as they do in real life. Users in VR scene cannot acquire any information about each other's gaze point. When they are discussing a phenomenon they are looking at, neither of them will know whether the other one is getting the wrong information, let alone give any correction or reminder. By visualizing the eye tracking data in VR, the other user's gaze information can be a obtained, which facilitates the efficiency of collaboration between two users and makes the interaction process as natural as in a real scene.

We proposed a technique with sharing real-time eye tracking data visualization between users in a collaborative VR environment. We built three different eye tracking data visualization modes as well as the no-eye-tracking modes, and compared the difference in the effectiveness of user collaboration with and without shared eye tracking data visualization. The contribution of this study is that we found gaze cursor that serves the best performance for the collaborative users improving efficiency and quality of collaborative work.

## 2   Related Work

Non-verbal cues, such as gaze, plays an important role in our daily communication, not only as a way of expressing intention, but also as a way of communicating information to the other. Oleg *et al.* [1] conducted a study on sharing visual attention between two players in a collaborative game so that one player's focusing area was visible to the other player. They investigated the difference between using head direction and eye gaze to estimate the point of attention, and the results showed that the duration of sharing eye gaze was shorter than sharing head direction, and the subjective ratings of teamwork were better in the high immersion condition. Wang *et al.* [2] investigated the use of gaze in a collaborative assembly task in which a user assembled an object with the assistance of a robot assistant. They found that being aware of a companion's gaze significantly improved collaboration efficiency. When gaze communication was available, task completion time was much shorter than when it was unavailable.

Newn *et al*. [3] tracked the user's gaze in strategic online games, and eye-based deception added difficulty and challenge to the game. D'Angelo *et al*. [4] designed novel gaze visualization for remote pair programming, and the programmers took more time to view the same code lines concurrently. They also designed gaze visualizations in remote collaboration to show collaborators where they were viewing in a shared visual space.

Various eye tracking devices have been used in single-player VR studies to accomplish different tasks. Kevin *et al*. [5] proposed a simulation of eye gaze in VR to improve the immersion of interaction between users and virtual non-player character (NPC). They developed an eye tracking interaction narrative system centered on the user's interaction with a gaze-aware avatar that responds to the player's gaze, simulating real human-to-human communication in a VR environment, and made preliminary measurements based on the user's responses. This study demonstrated that users had better experience during VR interactions with eye tracking. Boyd *et al*. [6] explored the effects of eye contact in immersive VR on children with autism. They developed an interaction system based on the eye tracking communication between the children and the avatars.

Visual attention prediction was crucial for predicting performance in motion. Heilmann *et al*. [7] investigated the difference between stimulus presentation and motor response in eye tracking studies, and examined the possibility of presenting this relationship in VR. Llanes-Jurado *et al*. [8] proposed a calibrated algorithm that can be applied to further experiments on eye tracking integrated into head-mounted displays and presented guidelines for calibrating the fixation point recognition algorithm.

## 3   Eye Tracking Date Visualization for Collaboration

### 3.1   Eye Tracking Method

We implemented eye tracking using a method based on pupil center corneal reflection (PCCR) [9]. An infrared camera was used to capture the user's eye image, after which the pupil center and the Purkinje were localized. The PCCR vector was calculated from the position of the pupil center and the location of the coordinates of the center of the Purkinje in the eye image [10].

The obtained PCCR feature vector was sent into the ray-tracing module in VR scene , hence  the radial of the feature vector is denoted as $X$, which represents the direction of the user's gaze. By re-establishing the local geometry and collision detection, we calculated and obtained the coordinates of the intersection point of the collision point $P$ and radial $X$.

### 3.2   Eye Tracking Data Visualization

In virtual reality environment, if user's visual attention behavior can be observed intuitively with eye tracking, it will be convenient for visual perception and cognition analysis in complex 3D scenes. Based on previous studies [11,12], we designed three kinds of eye tracking data visualization modes in virtual reality scenes: gaze cursor, gaze trajectory and gaze spotlight, as shown in Figure 1.

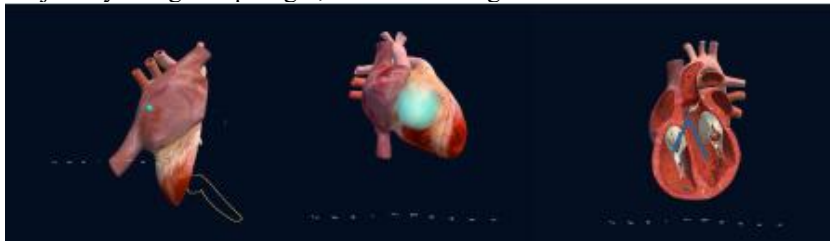

**Fig. 1.** Visualization of gaze points:  gaze cursor, gaze spotlight and gaze trajectory (from left to right).

**Gaze cursor :** is based on a blue sphere with a specific radius (*e,g*., radius is 5 in the world coordinate system). This visualization model is highly directional and has a clear, concise, and focused field of view [13].

**Gaze trajectory:** is to display the eye saccade, so that the original independent gaze points have a form of display with the chronological order.

**Gaze spotlight:** is with a range (radius is 40 in the world coordinate system) for local highlighting. This visual representation covers a larger area than gaze cursor, drawing attention to the around the gaze points.

### 3.3 Prototype System for Collaboration

In our research, we needed to build collaborative scenes and share eye tracking data between collaborators. The prototype system was developed using the Unity3D engine, which could access to external devices conveniently.

The system recorded and processed 3D scene data and the user's eye tracking data. Eye tracking modules, including infrared (IR) LEDs, IR lens, and high definition (HD) cameras, were added in the head mounted display (HMD) devices to track the user's eye tracking data [14] (as shown in Figure 2), and the high-precision eye tracking method we used in this study ensured that the accuracy was 0.5° (error was within the degree of visual angle). We used a framework based on the server-client model to synchronize simulations with the network. The users shared the same viewing angle in VR scene, which helped them to increase the sense of presence, eliminate motion sickness, and facilitate the collaboration between them.

## 4 Experiment

### 4.1 VR Scene

We designed 3D heart models in VR based on healthy heart and various diseased hearts, to simulate a controlled experiment of lesson of heart knowledge with different eye tracking data visualizations (gaze cursor, gaze trajectory, gaze spotlight). The user was allowed to learn about heart structure and disease in different eye tracking data visualization modes and no-eye tracking data visualization mode. After finishing the experiment, we compared which eye tracking data visualization mode was optimal in VR. The user was required to identify the heart model in VR and finish quiz about heart structure and disease. The answer time, , answer scores and eye tracking data were all recorded during their experiment.

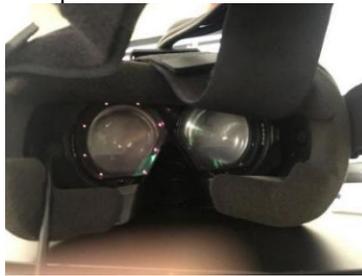

**Fig. 2.** VR HMD with eye tracking module in our study.

### 4.2 Experiment 1：Optimal Eye Tracking Data Visualization Modes

In this experiment, we invited a doctor (30 years old, female) from a local hospital. A VR HMD with an eye tracking module was used.

First, after calibrating the eye tracker, the doctor used VR HMD for a few minutes to familiarize with it, and then she gave a lecture about heart structure and related diseases in the VR environment. The lecture was recorded repeated and added with/without doctor's eye tracking data visualization.

**Participants:** We recruited 40 participants (26 males and 14 females, aged between 19 and 25) from the local participants pool and all participants were normal or corrected to normal vision, and they had no knowledge about heart structure and disease. 9 participants were familiar with VR and eye tracking. Before the experiment beginning, each participant signed an informed consent form and filled out a short background questionnaire.

**Groups:** Based on the different independent variables of eye tracking data visualizations, we divided the participants into 4 groups as follows (10 participants of each group). In addition, we added the speech of the lecture in different eye tracking data visualization modes as well as non-eye tracing data visualization mode. The doctor was required to teach the same content and keep similar visual attention behavior as much as possible for each lecture:

Group 1: gaze trajectory + speech;
Group 2: gaze spotlight + speech;
Group 3: gaze cursor + speech;
Group 4: no eye tracking data visualization + speech.

**Procedure:** Before the experiment began, participants were allowed to spend a few minutes familiarizing and adapting to the VR HMD.

Firstly, the participant was asked to learn from the doctor's teaching videos about the heart structure, mitral stenosis, aortic septal defect. Then the participant was required to wear the VR HMD and finish the quiz, which required the participant to use the handle to point out specific parts of the heart model (*i.e.*, the coronary artery, aorta, pulmonary artery, superior and inferior vena cava, left ventricle, right atrium, aortic valve, and mitral valve).

Secondly, the participant was asked to learn about the heart diseases (*e.g.*, symptoms caused by myocardial necrosis, mitral stenosis, aortic septal defect, atrial septal defect and ventricular septal defect) from the doctor's teaching videos with her eye tracking data visualizations. Then the participant was also required to wear the VR device to answer the quiz. The quiz would present the heart model with different heart diseases: atrial septal defect, mitral stenosis, ventricular septal defect and normal heart. Participants should select the correct name of the diseases accordingly. We recorded participants' score of answers, completion time and eye tracking data during the process of quiz answering in these two steps.

After completing the two quizzes, a questionnaire was used to collect participants' subjective feedback on VR eye tracking data visualization learning.

### 4.3 Experiment 2：Collaboration with Eye Tracking Data Visualizations

After we obtained the optimal eye tracking data visualization in Experiment 1, we conducted a collaborative work experiment on VR eye tracking data visualization.

**Participants:** we recruited 20 participants (12 males and 8 females, aged between 19 and 22) from the local participants pool and all participants were normal or corrected to normal vision and had no knowledge about heart structures and disease. No participants were familiar with VR and eye tracking. Each participant signed an informed consent form and filled out a short background questionnaire.

**Groups:** we randomly divided the participants into 2 groups:

Experimental group: with eye tracking data visualizations + speech (5 paired, and 10 participants totally);

Control group: without eye tracking data visualizations + speech (5 paired, and 10 participants totally);

**Procedure:** before the experiment began, each participant was allowed to spend a few minutes familiarizing and adapting to the VR HMD. Then each paired (two participants) were asked to wear VR HMDs, respectively, as shown in Figure 3.

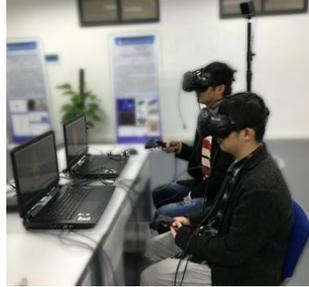

**Fig. 3.** Paired participants conducted collaborative work in VR.

In the VR environment, a diseased heart model was presented, and the participant needed to cooperate with the partner to recognize the diseases. Participants in the experimental group could observe each other's eye tracking data visualization, while the participants in the control group could not see eye tracking data visualization and only work together through free talking.

During the experiment, we recorded participants' eye tracking data in both experimental and control groups, and further analyzed the quality of collaboration based on participants' answers, communication records and eye tracking data.

## 5    Results

### 5.1    Optimal Eye Tracking Data Visualization

In order to explore the best eye tracking data visualization which accurately conveyed the partners' visual attention information as well as avoided causing excessive visual interference, we analyzed the results of Experiment 1.

It can be seen from the Table 1 that the correct rate of the heart structure and disease quiz of the gaze cursor was 76% and 82.5%, respectively, which was obviously superior to the other three visualization modes, while the average answer time of the heart structure of the gaze cursor (60.88s) was in the middle value of all the modes. We also found that for each experimental condition, the correct rate of the heart disease quiz was equal or higher than that of the heart structure quiz.

**Table 1.** Quiz results in different modes of eye tracking data visualization.

| Visualizations | Average correct rate of heart structure quiz | Average answer time of heart structure quiz | Average correct rate of heart disease quiz | Average answer time of heart disease quiz |
|---|---|---|---|---|
| Gaze trajectory | 40.00% | 56.62s | 40.00% | 124.284s |
| Gaze spotlight | 44.00% | 66.142s | 70.00% | 122.477s |
| Gaze cursor | 76.00% | 60.88s | 82.50% | 118.57s |
| No visualization | 44.00% | 59.25s | 60.00% | 99.529s |

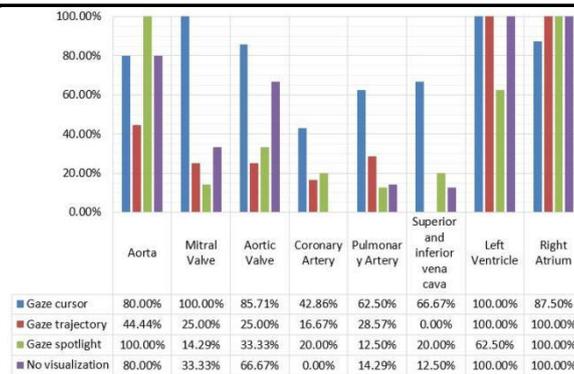

**Fig. 4.** Correctness rate for each question in each visualization in heart structure quiz.

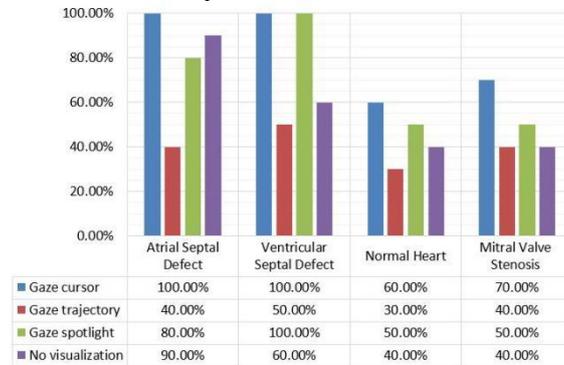

**Fig. 5.** Correctness rate for each question in each visualization in heart disease quiz.

Figure 4 summarized the correctness of each question in the heart structure quiz in the different modes of eye tracking data visualizations. We found that the correctness rate of aorta, left ventricle and right atrium were higher than the rest. This was because the aorta, left ventricle, and right atrium were more obvious and easy to be recognized in the heart structure distribution. In addition, we found that in the rest parts,, the gaze cursor showed a better performance than other eye tracking data visualizations. Although in aorta and right atrium, the gaze cursor mode did not outperform other modes, this was due to the errors in the calculation of the gaze point coordinates by the eye-tracking module, and the gaze cursor was small, which was more likely to cause the error.

Figure 5 summarizes of the correctness rate of each group of heart disease quiz in different modes of eye tracking data visualizations. We found that the correctness rates under the gaze trajectory mode were poor, even worse than the non-eye tracking data visualization. The reason was the gaze trajectory brought obvious visual interference to the participant, and distracted the participant's visual attention, so the participant could not obtain other's gaze data accurately. For example, in the case of ventricular septal defect and atrial septal defect, the area in the heart model were small but obvious. On the other hand, the gaze cursor and the gaze spotlight brought higher correctness rates, because these eye tracking data visualizations could accurately help participants find the diseased area.

We analyzed the correctness rates of each visualization modes by one-way ANOVA. In the heart structure quiz, we found a significant difference among the four eye tracking data visualization modes ($p<0.05$). Similarly, in the heart disease quiz, we also observed a significant difference ($p<0.001$) across the four modes, both results have proved that the quiz result has significant difference according to the different eye tracking data visualization modes under VR. However, we did not find significant difference at the answer time of heart structure quiz ($p=0.745$), or heart disease quiz ($p=0.428$), indicating the different eye tracking data visualization modes under VR did not affect participants to finish the quiz.

Therefore, we further conducted a *post-hoc* LSD test to compare the correctness rates of the gaze cursor mode and the rest three modes in heart structure quiz and heart disease quiz. As shown in Table 2, it can be seen that in the heart structure quiz, there was a significant difference between the gaze cursor and gaze trajectory ($p<0.001$), between gaze cursor and gaze spotlight ($p<0.005$), and between gaze cursor and no visualization mode ($p<0.005$). In the heart disease quiz, a significant difference could be observed between gaze cursor and gaze trajectory ($p<0.001$), between gaze cursor and no visualization mode ($p<0.05$), however, no significant difference could be observed between gaze cursor and gaze spotlight ($p=0.184$). These results suggested that the visualization modes of the gaze cursor was the optimal eye tracking data visualization mode under VR.

**Table 2.** Significant differences between gaze cursor and other eye tracking data visualization modes.

|  | Gaze trajectory | Gaze spotlight | No visualization |
|---|---|---|---|
| Heart structure quiz | $p<.001$ | $p<.005$ | $p<.005$ |
| Heart disease quiz | $p<.001$ | $p=.184$ | $p<.05$ |

### 5.2 Analysis of Collaborative Work

For the collaborative work, participants could observe the gaze cursor from the partners, and communicated more efficiently. We analyzed the eye tracking data of the participants during the collaboration, and compared their performance as well. Figure 6 showed the paired participants' eye tracking data in the gaze cursor mode and no visualization mode in time series. Different color and length of the bars indicated which part and how long did the participant spend to view in the heart model.

From Figure 6, 4 out of 5 groups using the eye tracking data visualizations successfully answered the quiz, while 2 out of 5 groups of participants who did not use the eye tracking data visualization successfully answered the quiz. In addition, it was obvious that the collaborative work using the gaze cursor spent less time than no visualization mode. This indicated that the shared eye tracking data visualization in VR could insist participants to finish the collaborative work more quickly.

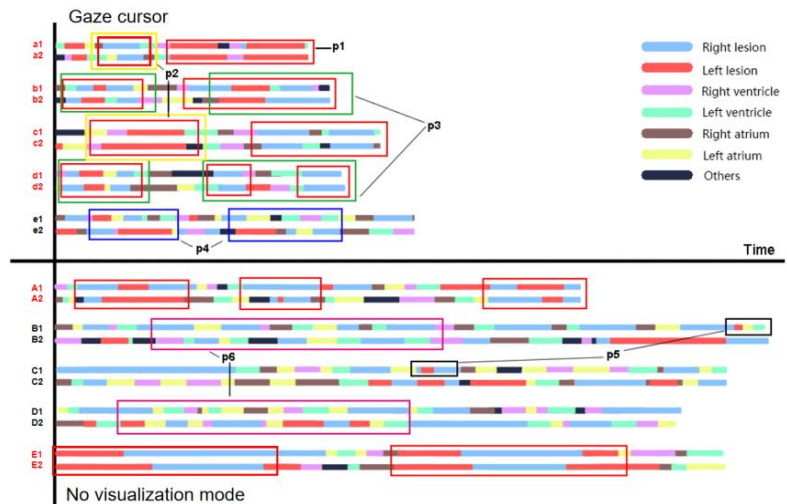

**Fig. 6.** 20 participants' eye tracking data between gaze cursor mode and no visualization mode Red box indicates the participants successfully answered the quiz.

Comparing the eye tracking data in the diseased area, we found that P1(highlighted with red box in the top of Figure 6) indicated the eye tracking data visualization helped the participants a1 and a2 successfully answered the quiz, and the eye tracking data distribution showed a highly coincidence, indicating that with the by the assistance of the gaze cursor mode, the paired participants could find and follow their partner's gaze. In no visualization mode, P6 (highlighted with the purple box in the bottom of Figure 6) indicated participants B1 and B2 (or D1 and D2) cluttered their gaze fixations, and they rarely discussed and communicated during the collaboration, thus they spent much time and did not answer the quiz correctly.

In the process of communication, when one of the participants using the gaze cursor wanted to discuss a certain area of the heart, it was easy and accurate to let the partner find out the exact area through the eye tracking. However, in no visualization mode, participants tent to spend longer time to let their partner understand and find the exact area, which was inefficient and time-wasting.

Furthermore, In the collaborative scenario, participants were found to positively observed their partner's eye tracking data visualization. The participants followed the orientation of partner's eye tracking data visualization and established an interaction based on eye tracking when working together. Interestingly, not all participants realized they had already found the diseased area at the first glance, only when they followed their partner's eye tracking data visualization did they realized that. For example, P2 (shown in the yellow box in the Figure 6), in the case of using the eye tracking data visualization, according to the speech records, we found that participant C2 who first saw the diseased area did not realize that that she found the exact area, but after the partner C1 followed C2's eye tracking, C1 successfully recognized the focus of the C2's gaze was on the diseased area, and reminded C2 immediately. In this process, they could collect information from each other' eye tracking data visualization and cooperate to complete the task.

In addition, it was suggested that paired participants could help mutually by looking at each other's eye tracking data visualization. For example, P3 (highlighted with the green box in the top of Figure 6) indicated participants d1 and d2 noticed the diseased area for the first time, but both of them were uncertain, then they resorted to each other's gaze , and discussed whether this was a diseased area. In this way, the eye tracking data visualization helped enhance the recognition of the diseased area for both participants.

In the no-eye tracking data visualization mode, for example, P5 (highlighted with the black box in the bottom of Figure 6) indicated that without eye tracking data visualization, for participants B1 and B2 (or C1 and C2), one observed the diseased area, while the other didn't. In this case, participants were prone to be confused. For example, one participant said in the post experiment interview: "*I did not know whether what I saw was the diseased area; I'm not sure if I need to communicate with my partner, so I spent much time and did not complete the task at last.*"

Eye tracking data visualization also required the participants seriously to think and discuss carefully. P4 (highlighted with the blue box in the last line of Figure 6) indicated both participant e1 and e2 had seen the diseased area, but they were all distracted by each other's eye tracking data visualization. They did not discuss what they observed in time, only blindly followed each other's gaze. It wasted much time and led to the failure of the task.

## 6    Conclusion

The complex VR environment and model structure bring lots of challenges for users to accomplish collaborative tasks, and they cannot communicate and conduct collaborative work with them like in the real world. This study designed eye tracking data visualization and utilized the visualization as visual attention indicators for paired users during their collaboration. We fund that gaze cursor was the best visualization modes, and applied it to facilitate the collaborative work in the heart lecture scene, and the experimental results showed that it could improve the quality and efficiency of the collaboration in VR environment.

# 7 Acknowledgement

The authors would like to thank all the volunteers who participated in the experiments. This work was supported in part by the National Natural Science Foundation of China under Grant 62172368, 61772468, and the Natural Science Foundation of Zhejiang Province under Grant LR22F020003.